\documentclass[10pt,a4paper]{article}
\usepackage{amsmath,amssymb,latexsym}
\expandafter\ifx\csname amssym.def\endcsname\relax \else\endinput\fi
\expandafter\edef\csname amssym.def\endcsname{%
       \catcode`\noexpand\@=\the\catcode`\@\space}
\catcode`\@=11
\def\undefine#1{\let#1\undefined}
\def\newsymbol#1#2#3#4#5{\let\next@\relax
 \ifnum#2=\@ne\let\next@\msafam@\else
 \ifnum#2=\tw@\let\next@\msbfam@\fi\fi
 \mathchardef#1="#3\next@#4#5}
\def\mathhexbox@#1#2#3{\relax
 \ifmmode\mathpalette{}{\m@th\mathchar"#1#2#3}%
 \else\leavevmode\hbox{$\m@th\mathchar"#1#2#3$}\fi}
\def\hexnumber@#1{\ifcase#1 0\or 1\or 2\or 3\or 4\or 5\or 6\or 7\or 8\or
 9\or A\or B\or C\or D\or E\or F\fi}

\font\tenmsa=msam10
\font\sevenmsa=msam7
\font\fivemsa=msam5
\newfam\msafam
\textfont\msafam=\tenmsa
\scriptfont\msafam=\sevenmsa
\scriptscriptfont\msafam=\fivemsa
\edef\msafam@{\hexnumber@\msafam}
\mathchardef\dabar@"0\msafam@39
\def\dashrightarrow{\mathrel{\dabar@\dabar@\mathchar"0\msafam@4B}}
\def\dashleftarrow{\mathrel{\mathchar"0\msafam@4C\dabar@\dabar@}}

\def\ulcorner{\delimiter"4\msafam@70\msafam@70 }
\def\urcorner{\delimiter"5\msafam@71\msafam@71 }
\def\llcorner{\delimiter"4\msafam@78\msafam@78 }
\def\lrcorner{\delimiter"5\msafam@79\msafam@79 }
\def\yen{{\mathhexbox@\msafam@55}}
\def\checkmark{{\mathhexbox@\msafam@58}}
\def\circledR{{\mathhexbox@\msafam@72}}
\def\maltese{{\mathhexbox@\msafam@7A}}
\def\circledS{{\mathhexbox@\msafam@73}}

\font\tenmsb=msbm10
\font\sevenmsb=msbm7
\font\fivemsb=msbm5
\newfam\msbfam
\textfont\msbfam=\tenmsb
\scriptfont\msbfam=\sevenmsb
\scriptscriptfont\msbfam=\fivemsb
\edef\msbfam@{\hexnumber@\msbfam}
\def\Bbb#1{{\fam\msbfam\relax#1}}
\def\widehat#1{\setbox\z@\hbox{$\m@th#1$}%
 \ifdim\wd\z@>\tw@ em\mathaccent"0\msbfam@5B{#1}%
 \else\mathaccent"0362{#1}\fi}
\def\widetilde#1{\setbox\z@\hbox{$\m@th#1$}%
 \ifdim\wd\z@>\tw@ em\mathaccent"0\msbfam@5D{#1}%
 \else\mathaccent"0365{#1}\fi}
\font\teneufm=eufm10
\font\seveneufm=eufm7
\font\fiveeufm=eufm5
\newfam\eufmfam
\textfont\eufmfam=\teneufm
\scriptfont\eufmfam=\seveneufm
\scriptscriptfont\eufmfam=\fiveeufm
\def\frak#1{{\fam\eufmfam\relax#1}}

\csname amssym.def\endcsname

\makeatletter
\def\section{\@startsection {section}{1}{\z@}{-3.5ex plus -1ex minus
 -.2ex}{2.3ex plus .2ex}{\large\sc}}
\def\subsection{\@startsection{subsection}{2}{\z@}{-3.25ex plus -1ex minus
 -.2ex}{1.5ex plus .2ex}{\normalsize\sc}}
\makeatother

\newcommand{\nc}{\newcommand}
\newcommand{\rnc}{\renewcommand}

\nc{\subs}[1]{{\vspace*{0.5cm}}%
{\noindent\underline{\small\sc #1}}{\addcontentsline{toc}{subsubsection}{#1}}%
{\vspace*{0.3cm}}}

\nc{\subss}[1]{{\vspace*{0.5cm}}%
{\noindent\underline{\small\sc #1}}%
{\vspace*{0.3cm}}}

\nc{\chap}[1]{{\clearpage}%
\begin{center}%
{\noindent\underline{\large\sc #1}}{\addcontentsline{toc}{section}{#1}}%
\end{center}%
{\vspace*{0.3cm}}}

\nc{\be}{\begin{equation}}
\nc{\ee}{\end{equation}}
\nc{\bea}{\begin{eqnarray}}
\nc{\eea}{\end{eqnarray}}

\nc{\trac}[2]{{\textstyle\frac{#1}{#2}}}

\nc{\ex}[1]{\mbox{e}^{\,\textstyle#1}}

\nc{\CC}{\Bbb{C}}
\nc{\HH}{\Bbb{H}}
\nc{\PP}{\Bbb{P}}
\nc{\RR}{\Bbb{R}}
\nc{\ZZ}{\Bbb{Z}}
\nc{\II}{\Bbb{I}}
\nc{\EE}{\Bbb{E}}
\nc{\TT}{\Bbb{T}}
\nc{\DD}{\mathrm{I}\!\mathrm{D}}

\rnc{\a}{\alpha}
\rnc{\b}{\beta}
\rnc{\d}{\delta}
\nc{\ga}{\gamma}
\nc{\la}{\lambda}
\nc{\f}{\phi}
\nc{\p}{\psi}
\nc{\e}{\eta}
\rnc{\c}{\chi}
\nc{\eps}{\epsilon}
\nc{\om}{\omega}
\nc{\Om}{\Omega}

\nc{\symx}{\circledS}
\newsymbol\smallsmile 1360
\newsymbol\smallfrown 1361
\nc{\ad}{\mathop{\mbox{ad}}\nolimits}
\nc{\tr}{\mathop{\mbox{tr}}\nolimits}
\nc{\Tr}{\mathop{\mbox{Tr}}\nolimits}
\nc{\Det}{\mathop{\mbox{Det}}\nolimits}
\rnc{\det}{\mathop{\mbox{det}}\nolimits}
\nc{\rk}{\mathop{\mbox{rk}}\nolimits}
\nc{\del}{\partial}
\nc{\diag}{\mathop{\mbox{diag}}\nolimits}
\nc{\ra}{\rightarrow}
\nc{\Ra}{\Rightarrow}
\nc{\LRa}{\Leftrightarrow}
\nc{\lra}{\leftrightarrow}
\nc{\ot}{\otimes}
\rnc{\ss}{\subset}
\nc{\nul}{\noindent\underline}
\nc{\non}{\nonumber\\}
\nc{\mat}[4]{\left(\begin{array}{cc}#1&#2\\#3&#4\end{array}\right)}
\rnc{\lg}{\frak{g}}
\nc{\G}[3]{\Gamma^{#1}_{\;{#2}{#3}}}
\nc{\nam}{\nabla_{\mu}}
\nc{\nan}{\nabla_{\nu}}
\nc{\dx}{\dot{x}}
\nc{\dxl}{\dot{x}^{\la}}
\nc{\dxm}{\dot{x}^{\mu}}
\nc{\dxn}{\dot{x}^{\nu}}
\nc{\ddx}{\ddot{x}}
\nc{\ddxm}{\ddot{x}^{\mu}}
\nc{\ddxn}{\ddot{x}^{\nu}}
\nc{\dxi}{\dot{\xi}}
\nc{\ddxi}{\ddot{\xi}}

\def \p {\phi}

\begin{document}


\begin{center}
{\Large\sc Brown-York Energy and Radial Geodesics}
\end{center}
\vspace{0.2cm}

\begin{center}
{\large\sc Matthias Blau} and {\large\sc Blaise Rollier}
\end{center}

\vskip 0.2 cm
\centerline{\it Institut de Physique, Universit\'e de Neuch\^atel}
\centerline{\it Rue Breguet 1, CH-2000 Neuch\^atel, Switzerland} 

\vspace{1cm}

We compare the Brown-York (BY) and the standard Misner-Sharp (MS)
quasilocal energies for round spheres in spherically symmetric space-times
from the point of view of radial geodesics. In particular, we show that
the relation between the BY and MS energies is precisely analogous to that
between the (relativistic) energy $E$ of a geodesic and the effective
(Newtonian) energy $E_{\text{eff}}$ 
appearing in the geodesic equation, thus
shedding some light on the relation between the two. Moreover, for
Schwarzschild-like metrics we establish a general relationship between
the BY energy and the geodesic effective potential which explains and
generalises the recently observed connection between negative BY energy
and the repulsive behaviour of geodesics in the Reissner-Nordstr\o m
metric. We also comment on the extension of this connection between
geodesics and the quasilocal BY energy to regions inside a horizon.





\section{Introduction}

It is a consequence of the fundamental general covariance of general
relativity that there is no well-defined covariant notion of the local
energy density of the gravitational field. The next best thing is perhaps
the notion of a quasilocal energy (QLE), i.e.\ the energy contained
in a two-dimensional surface.  Numerous definitions of QLE have been
proposed in the literature (for a detailed and up-to-date review with many
references see \cite{szab}), and these tend to be mutually inequivalent
even in simple cases such as the Kerr metric \cite{berg}.

There is at least one case, however, in which there appears to be
\textit{almost} universal agreement as to what the QLE should be,
namely for round spheres (i.e.\ orbits of the rotational isometry
group) in spherically symmetric space-times. In that case, the
classical Misner-Sharp (MS) energy \cite{ms} (see e.g.\ \cite{szab} or
\cite{hayward} for recent discussions) is widely considered to be the
``standard'' definition of the energy for round spheres.

One serious contender to this definition is based on the Brown-York (BY)
QLE \cite{by1}. The definition of the BY energy is based on the covariant
Hamilton-Jacobi formulation of general relativity, and this makes it a
natural object to consider in a variety of contexts, with numerous 
attractive features. However, the standard
BY energy for round spheres does not 
agree with the standard MS energy (even for the Schwarzschild metric), and
this fact has occasionally been used as an argument against the BY energy
as a ``good'' definition of a QLE (see e.g.\ the discussions in
\cite{szab,hayward}).

In this article we will look at the relationship and differences between
the MS and BY energies for round spheres from the point of view of
geodesics and
their associated energy concepts like the relativistic geodesic energy
and the effective Newtonian potential. In general, one would not expect
point-like objects to be able to probe something not quite local like a
QLE. However, the situation is different for round spheres for which the
QLE is independent of the angular coordinates. In such a situation it is
fair to ask whether there is a relation between the gravitational energy
as felt by a point-like observer (geodesic) and that defined according
to some QLE prescription.

Originally, our investigation of these issues was prompted by an
observation and a remark in \cite{lsy}. There it was observed that
for the Reissner-Nordstr\o m metric the BY energy becomes negative
for sufficiently small radius. In \cite{lsy} it was suggested that
this negative energy is strictly related to the well-known repulsive
behaviour exhibited by the geodesics of massive neutral particles in
the Reissner-Nordstr\o m metric.  

What supports this point of
view is the fact that the energy indeed becomes negative at precisely
the radius where radial geodesics begin to experience the repulsive
behaviour of the Reissner-Nordstr\o m core. This clearly hints at a
deeper connection between geodesic and quasilocal energy, or, in the words of
\cite{lsy}:``\textit{The turnaround radius agrees with the radius where
the quasilocal energy becomes negative, so it seems that the two effects
are very likely connected.}''
We will indeed be able to establish a general relationship between the BY
energy and the geodesic effective potential (for radial geodesics) and,
in particular, a relation between negative BY energy and a repulsive
behaviour for geodesics. 

Our results also shed some light on the difference between the MS and
BY energies for round spheres.  In particular, for Schwarzschild-like
metrics $ds^2 = -f(r)^2 dt^2 + f(r)^{-2} dr^2 + r^2 d\Omega^2$, for
which geodesics are conveniently described in terms of an effective
Newtonian potential, we observe that the MS energy is directly related
to the effective potential $V_{\text{eff}}(r)$ for radial goedesics, and that
the relation between the MS and BY energies is strictly analogous to the
relation $E_{\text{eff}} = \trac{1}{2}(E^2 -1)$ between the energy appearing in
the effective potential equation and the relativistic geodesic energy $E$
of the particle.  Therefore, inasmuch as $E$ is a relativistic energy and
$E_{\text{eff}}$ an effective Newtonian energy, perhaps one interpretation of
the difference between the BY and MS energies for round spheres is to say
that the former provides one with a relativistic notion of gravitational
energy while the MS energy is more like an effective Newtonian quantity.

We also briefly discuss the extension of these results to regions inside
a horizon. An extension of the BY energy to this case was proposed in
\cite{lsy}. However, it has been remarked\footnote{e.g.\ by one of the
referees, and by Ruth Durrer (private communication)} that the proposal of
\cite{lsy} should perhaps better be thought of as a quasi-local momentum.
Our geodesic perspective is compatible with this
point of view since, as we will show, the relation between the BY
``energy'' of \cite{lsy} and the effective potential inside the horizon
is identical to that between $E$ and $E_{\text{eff}}$ provided that one
considers geodesics that are \textit{spacelike} (outside the horizon).

We believe that the message of this work is two-fold: First of all,
it shows that there are situations where geodesic test particles can be
useful to probe candidate definitions of QLE.  Moreover, these results
also illuminate the difference between the BY and MS energies
and provide further evidence that the BY definition of a QLE provides
a good (relativistic) measure of the gravitational energy even though
(or even precisely because) it does not agree with the standard (and
perhaps somewhat more Newtonian) MS energy for round spheres.

\section{Brown-York and Misner-Sharp Energy for Spherical Symmetry}

We briefly recall the definition of the BY and MS energies
for round spheres, referring to the original literature (e.g.\
\cite{by1,by2,lsy} and \cite{ms}) 
and the review article \cite{szab} for details.
We will consider a general
spherically symmetric metric written in the form 
\be
ds^2 = -N(t,r)^2 dt^2 + f(t,r)^{-2}dr^2 + r^2 d\Omega^2
\label{met2}
\ee
with $d\Omega^2= d\theta^2 + \sin^2\theta d\phi^2$ 
the standard line-element on the unit 2-sphere. 
Even though we will
only consider 4-dimensional space-times in this article, the extension to
higher dimensions is rather straightforward. 
In contrast to \cite{by2} we prefer to work directly
with the area radius $r$ as the radial coordinate. 
The \textit{round spheres} in this space-time (the orbits of the
rotational isometry group) are the 2-spheres $t=\text{const.},
r=\text{const}$.
It was shown in \cite{by1,by2} that, 
in any region in which $\del_t$ is timelike and $\del_r$ is
spacelike, the standard 
BY quasilocal energy $E_{BY}(t,r)$ associated to a round sphere
of radius $r$, and calculated with respect to the standard static
observers associated to the spatial slicing $t=\text{const.}$
is given by
\be
E_{BY}(t,r) = \frac{r}{G_N}(1-f(t,r))\;\;,
\label{eby2}
\ee
where $G_N$ is Newton's constant. 

This BY energy differs from the ``standard'' 
Misner-Sharp (MS) energy \cite{ms} for round spheres which, for a metric of 
the type (\ref{met2}) and for any $(t,r)$, is given by
\be
E_{MS}(t,r) = \frac{r}{2G_N}\left(1-f(t,r)^2\right)\;\;.
\label{ems1}
\ee
For example, for the Reissner-Nordstr\o m metric
$N(r)^2 = f(r)^2 = 1 - \frac{2m}{r}  + \frac{e^2}{r^2}$
one has
\be
\begin{aligned}
E_{BY}(r) &= \frac{r}{G_N}
\left(1-\sqrt{1 - \frac{2m}{r}+\frac{e^2}{r^2}}\right)\\
E_{MS}(r) &= \frac{1}{G_N}\left(m-\frac{e^2}{2r}\right)\;\;.
\end{aligned}
\ee
Both reduce to the ADM mass $M=m/G_N$ asymptotically,
$\lim_{r\ra\infty} E_{BY}(r) =
\lim_{r\ra\infty} E_{MS}(r) = M$
(and for the Schwarzschild metric one evidently
has $E_{MS}(r)=M$ for all $r$).
Moreover, for sufficiently small values of $r$, $r<r_0 = e^2/2m$ 
both the MS and the BY energy are negative (note that the expression for the
BY energy is also valid inside the inner horizon $r_-$ and that $r_0 < r_-$).
The qualitative (and not just quantitative) difference bewteen the
BY and MS energies is e.g.\ illustrated by the fact that, 
unlike the MS energy, the  BY energy is finite at $r=0$,
$E_{BY}(0)=-|e|/G_N$ \cite{lsy}.

\section{Brown-York and geodesic energy for Schwarzschild-like metrics}

In order to analyse the BY energy (and its relationship with the MS energy)
from the point of view of geodesics, we now specialise to Schwarzschild-like
metrics, i.e.\ static spherically symmetric metrics with $N(r)=f(r)$ (the
extension to time-dependent Schwarzschild-like metrics with $f=f(t,r)$ is 
straightforward),
\be
ds^2 = -f(r)^2 dt^2 + f(r)^{-2} dr^2 + r^2 d\Omega^2\;\;.
\label{met4}
\ee
In this case the behaviour of timelike radial 
geodesics is governed by the effective potential equation 
\be
\trac{1}{2}\dot{r}^2 + V_{\text{eff}}(r) = E_{\text{eff}}\;\;,
\label{veff2}
\ee
where the effective (and effectively Newtonian) potential
$V_{\text{eff}}(r)$ is related to $f(r)^2$ by
\be
f(r)^2 = 1 + 2 V_{\text{eff}}(r)\;\;,
\label{fv}
\ee
and the effective energy $E_{\text{eff}}$ is given in terms of the 
relativistic geodesic energy per unit rest mass 
$E=f(r)^2 \dot{t}$ of the particle by
\be
E_{\text{eff}} = \trac{1}{2}(E^2 - 1)\;\;.
\label{eeff}
\ee
For later we note that $E=f(r_m)$ where $r_m$ (the index could indicate 
a minimum or maximum) is a turning point, $\dot{r}_m=0$, of the trajectory,
and that 
in the asymptotically flat case (which we take here to simply mean
$\lim_{r\ra\infty} f(r)=1$) for scattering trajectories that reach 
(or start out at) $r\ra\infty$ one also has the relation 
$E^2 = 1 + \dot{r}_{\infty}^2\geq 1$ between $E$ and 
the velocity at infinity. 
In particular, for scattering trajectories in 
the Reissner-Nordstr\o m metric, for the minimal radius $r_m=r_m(E)$ one has
$r_m(E) \leq r_m(E=1) = e^2/2m$, 
which, as noted in \cite{lsy}, agrees with the radius 
$r_0$ where the BY (and MS) energy becomes negative.  

In order to now
study the relations among the Brown-York energy, the Misner-Sharp
energy, and the geodesic effective potential, it turns out to be convenient to 
introduce the corresponding potentials
\be
V_{BY}(r) := -G_{N}\frac{E_{BY}(r)}{r}\qquad\qquad
V_{MS}(r) := -G_{N}\frac{E_{MS}(r)}{r}\;\;.
\ee
Using the definition (\ref{ems1}) of the MS energy and (\ref{fv}),
one immediately sees that
\be
V_{MS}(r) = -\trac{1}{2}(1-f(r)^2) = V_{\text{eff}}(r)\;\;.
\label{mseff}
\ee
Thus the MS potential agrees on the nose with the effective potential
and has a clear physical interpretation in the present context. In 
particular, negative MS energy is strictly correlated with a repulsive
behaviour of the effective potential for radial geodesics.

What about the BY potential? Given the above relation between the
MS energy and radial geodesics, one's first thought may perhaps
be\footnote{This was not our first thought, but we are grateful to one
of the referees for reminding us that it should perhaps have been.}
that to establish a link with the BY energy one should calculate the
latter for freely falling (geodesic) rather than static observers,
or for static observers in comoving (Novikov) coordinates.
The Schwarzschild BY energy for geodesic obervers was first determined
in \cite{bm} and more recently, also motivated by the appearance of
the first version of the present article on the arXiv, in \cite{yc}
(with a slightly different prescription).
For example, the result of \cite{bm} (for an observer initially at rest at
infinity) is
\be
E_{BY}^{\text{freefall}}(r) = \frac{r}{G_N}
\left(\sqrt{1+ \frac{2m}{r}}-1\right)\;\;.
\ee
In Novikov coordinates $(\tau,R)$, where $\tau$ is the proper time of a
radially infalling observer and $R$ is related to the maximal radius 
$r_m$ of the geodesic by $R= \sqrt{\frac{r_m}{2m}-1}$,
the Schwarzschild metric reads (see e.g.\ \cite[\S 31.4]{mtw})
\be
ds^2 = -d\tau^2 + \frac{R^2+1}{R^2} \left(\frac{\del r}{\del R}\right)^2 dR^2
+ r(\tau,R)^2 d\Omega^2\;\;.
\ee
Calculating the BY energy for ``static'' observers in this space-time, one
finds 
\be
\label{enov}
E_{BY}^{\text{Novikov}}(r) = \frac{r}{G_N}
\left(1 - \frac{R}{\sqrt{R^2 +1}}\right)
= \frac{r}{G_N}
\left(1 - \sqrt{1-\frac{2m}{r_m}}\right)\;\;.
\ee
This can e.g.\ be seen to agree with the result of
\cite{yc}, based on the calculation of the freefall BY energy in Kruskal
coordinates. Thus, neither do the above results reproduce the MS
energy, nor do they appear to be related to the effective geodesic potential
in any other particularly useful or illuminating way.

In this context it is perhaps also worth pointing out that in
\cite{by2} a change of coordinates (foliation) $t\ra T(t,r)$ for the
Schwarzschild metric was exhibited with respect to which the standard
BY energy takes the MS value $E_{BY}(r) = M$. Such a foliation, giving
$E_{BY}(r)=E_{MS}(r)$, can also readily be constructed for the general
Schwarzschild-like metric \eqref{met4}. It suffices to choose $T(t,r)$ such
that
\be
\label{T}
dT = dt + \frac{1-f^2}{f^2(1+f^2)} dr
\;\;.
\ee 
To see this note that for a general spherically symmetric metric of the
form 
\be
ds^2 = -N(t,r)^2 dt^2 + F(t,r)^{-2}(dr + A(t,r)dt)^2 + r^2 d\Omega^2
\ee
the BY energy is still given by \eqref{eby2} (with $f\ra F$), 
and that with the choice \eqref{T} one has
$1-F = \trac{1}{2}(1-f^2)$, so that indeed $E_{BY}(r) = E_{MS}(r)$.
However, the physical significance of this choice of foliation escapes us, and
this construction does not appear to shed any light on the relationship
between the BY energy and geodesic notions of energy.

Thus we now return to the task of relating the standard BY energy
\eqref{eby2} to the geodesic effective potential.
Substituting (\ref{fv}) in (\ref{eby2}), one finds
\be
E_{BY}(r) = \frac{r}{G_N}(1-\sqrt{1+2 V_{\text{eff}}(r)})\;\;,
\ee
or
\be
1+ V_{BY}(r) = \sqrt{1+2V_{\text{eff}}(r)}\;\;.
\label{vbyveff}
\ee
While this relation, which we may also read as the relation between the MS
energy and the BY energy, may appear to be somewhat obscure, it reveals several
interesting features of the BY potential $V_{BY}(r)$
and its relation to $V_{\text{eff}}(r)=V_{MS}(r)$:

\begin{enumerate}

\item First of all 
we observe that (\ref{vbyveff}) and (\ref{veff2}) 
allow us to express the BY potential in terms of geodesic quantities as
\be
\label{vbyr}
1+ V_{BY}(r) = \sqrt{E^2 -\dot{r}^2}\leq E\;\;.
\ee
In other words, the relation between the Brown-York potential and the 
relativistic energy $E$ 
(per unit rest mass) of the particle can be phrased as 

\textit{The energy $E$ of the geodesic particle is greater or equal to 
the sum of its rest mass and the gravitational potential energy (as measured
by $V_{BY}(r)$), with equality at points where $\dot{r}=0$.}

Thus the BY potential appears to provides a reasonable
measure of the energy of the gravitational field in this context.
The inequality $E\geq 1+ V_{BY}(r)$ should be compared and
contrasted with the analogous equation $E_{\text{eff}} \geq V_{MS}(r)$
for the MS (or effective) potential that follows from (\ref{veff2}).
This suggests a certain
 analogy $E_{BY} \lra E$ and $E_{MS} \lra E_{\text{eff}}$.

\item This analogy is strengthened by the observation that \eqref{vbyveff}
implies
\be
V_{\text{eff}}(r) = \trac{1}{2}\left((1+V_{BY}(r))^2-1\right),
\label{vv2}
\ee
which shows that the relation between $V_{\text{eff}}$ and 
$1+V_{BY}$ is identical to the relation $E_{\text{eff}} = \frac{1}{2}(E^2-1)$
\eqref{eeff}
between the effective energy $E_{\text{eff}}$ and the geodesic particle
energy $E$.

Thus, since $E$ is a relativistic energy and $E_{\text{eff}}$ an effective
Newtonian quantity, it is tempting to say that the BY energy provides one
with a relativistic notion of gravitational energy while the MS energy
is really more like an effective Newtonian quantity. So far, however,
this is only a suggestion, based on the geodesic analogy that we have
developed here, and further analysis of this issue, in other settings,
will be required to substantiate (or disprove) this interpretation of
the difference between $E_{MS}$ and $E_{BY}$.

\item
Finally, \eqref{vbyveff} implies that
$V_{\text{eff}}(r)$ and 
$V_{BY}(r)$ have the same zeros
and that the BY potential is repulsive/positive whenever (and
whereever) the effective potential is repulsive.
Thus the BY energy is negative if and only if the effective
potential is repulsive. In particular, \eqref{vbyr} leads
to a simple expression for the BY energy at any
turning point $r_m$ ($\dot{r}_m=0$) of the potential, namely 
$1+ V_{BY}(r_m) = E$ or 
\be
E_{BY}(r_m) = \frac{r_m}{G_N}(1-E)\;\;.
\ee
This also follows directly from the definition \eqref{eby2} and the
previously noted $E=f(r_m)$.
In particular, $E_{BY}(r_m)$ is negative for scattering trajectories with
$E> 1$. Thus non-positive BY energy is necessary for a repulsive 
behaviour of radial geodesics. This provides a simple explanation and
proof of a generalisation of the 
observation made in \cite{lsy} in the context of the Reissner-Nordstr\o m
metric. Note also that, for the Schwarzschild metric, at $r=r_m$ one has
$E_{BY}(r_m)=E_{BY}^{\text{Novikov}}(r_m)$, 
so that static and freely falling
observers can agree on the energy at a turning point of the freely falling
observer, as they should.
\end{enumerate}

All in all this provides us with a coherent picture of the relation
between geodesic notions of energy on the one hand, and the quasilocal
gravitational MS and BY energies for round spheres on the other.

Finally we comment briefly, from the present geodesic point of view,
on the extension $E_{LSY}(r)$ 
of the BY energy $E_{BY}(r)$ to the interior of a horizon proposed in 
\cite{lsy}. Writing the Schwarzschild-like metric as
\be
ds^2 = -\epsilon f(r)^2 dt^2 + \epsilon f(r)^{-2} dr^2 + r^2 d\Omega^2\;\;,
\label{met5}
\ee
with $\epsilon=\pm 1$ corresponding to the exterior (interior) region, 
the definition of \cite{lsy} is
\be
E_{LSY}(r) = \frac{r}{G_N}(1-\epsilon f(r))
\ee
($E_{LSY}(r) = E_{BY}(r)$ in the region $\epsilon=+1$).
We now write the effective potential equation for radial timelike
($\lambda=+1$) or spacelike ($\lambda=-1$) geodesics as
\be
\trac{1}{2}\dot{r}^2 + V_{\text{eff}}^{\lambda}(r) = E_{\text{eff}}^{\lambda}
\ee
where
$E_{\text{eff}}^{\lambda} = \trac{1}{2}(E^2-\lambda)$.
Then one easily finds 
\be
\epsilon\lambda V_{\text{eff}}^{\lambda}(r) = 
\trac{1}{2}\left((1+V_{LSY}(r))^2-\epsilon\right)\;\;.
\label{vv3}
\ee
This relation between the effective potential
$V_{\text{eff}}^{\lambda}(r)$ and 
$1+V_{LSY}(r)$ is identical to the relation 
between the effective Newtonian energy $E_{\text{eff}}^\lambda$
and the relativistic geodesic 
energy $E$ for any $\epsilon$ provided that one 
correlates the region of interest (specified by
$\epsilon$) with the character of the geodesic (indicated by $\lambda$)
by making the choice
$\epsilon = \lambda$.
Thus for $\epsilon=-1$ 
$E_{LSY}(r)$ appears to be naturally associated with spacelike
geodesics. 

\subsection*{Acknowledgements}

We are grateful to the referees for their useful comments and suggestions.
This work forms part of a Master thesis project of B.R.\ performed jointly
at the Universit\'e de Gen\`eve and the Universit\'e de Neuch\^atel. M.B.\
acknowledges financial support from the Swiss National Science Foundation
and the EU under contract MRTN-CT-2004-005104.

\rnc{\Large}{\normalsize}

\end{document}